\documentclass[11pt]{article}
\pdfoutput=1
\usepackage{graphicx}
\usepackage{epstopdf} 
\usepackage{epsfig}
\usepackage{amsmath}
\usepackage{amssymb}
  \usepackage{color}
  \usepackage{verbatim}
\usepackage{multirow}
\usepackage[bottom]{footmisc}
\usepackage{cite,url}
\usepackage{comment}

\usepackage{booktabs,multirow,tabularx}

\newcommand{\sineff}{\mbox{$\sin^2 \theta^{{\rm lep}}_{{\rm eff}}$} }
\newcommand\php{\phi^+}
\newcommand\phm{\phi^-}
\newcommand\phu{\phi_1}
\newcommand\phd{\phi_2}
\newcommand\Phid{\Phi^\dagger}

\newcommand{\drcarw}{\mbox{$\Delta \hat{r}_{W}$}}

\newcommand{\drcarwd}{\mbox{$\Delta \hat{r}^{(2)}_{W}$}}

\newcommand{\bms}{\mbox{$\overline{MS}$}}
\newcommand{\ms}{\mbox{$\overline{\scriptstyle MS}$}}
\newcommand{\sincur}{\mbox{$\sin^{2}\!\hat{\theta}_{W}$}}

\newcommand{\scs}{\mbox{$\hat{s}^{2}$}}

\newcommand{\ccs}{\mbox{$\hat{c}^{2}$}}
\newcommand{\alc}{\mbox{$\hat{\alpha}$}}
\newcommand{\kcar}{\hat{k}}
\newcommand{\rhoh}{\mbox{$\hat{\rho}$}}
\newcommand{\aww}{A_{\scriptscriptstyle WW}}
\newcommand{\azz}{A_{\scriptscriptstyle ZZ}}

\newcommand{\nn}{\nonumber}

\newcommand\sss{\scriptscriptstyle}

\newcommand{\tril}{\lambda_{3}}
\newcommand{\trilsm}{\tril^{\rm SM}}
\newcommand{\qual}{\lambda_{4}}
\newcommand{\mh}{m_{ \sss H}}
\newcommand{\mw}{m_{ \sss W}}
\newcommand{\mz}{m_{ \sss Z}}

\newcommand{\mhw}{\zeta_{\scriptscriptstyle W}}

\newcommand{\lnb}{\mbox{$\overline{\mbox{ln}}$}}

\newcommand{\ktre}{\kappa_{\lambda}}

\newcommand{\ggF}{gg{\rm F}}
\def\beq{\begin{equation}}
\def\beqn{\begin{eqnarray}}
\def\eeq{\end{equation}}
\def\eeqn{\end{eqnarray}}
\def\beal{\begin{align}}
\def\endal{\end{align}}

\newenvironment{appendletterA}
 {
  \typeout{ Starting Appendix \thesection }
  \setcounter{section}{0}
  \setcounter{equation}{0}
  
 }{
  \typeout{Appendix done}
 }

\vspace*{-1cm}
\begin{document}\hfill
{\flushright{
        \begin{minipage}{3.0cm}
          RM3-TH/17-1
        \end{minipage}        }

}
\vspace*{1cm}
\color{black}
\begin{center}
{\Large \bf  \color{blue} Constraints on the trilinear Higgs 
 self coupling from
 precision observables}

\bigskip\color{black}\vspace{.4cm}
{\large\bf G.~Degrassi$^a$, M.~Fedele$^b$, P.P.~Giardino$^{c}$}
\\[7mm]
{\it  (a) Dipartimento di Matematica e Fisica, Universit{\`a} di Roma Tre and \\
 INFN, sezione di Roma Tre, I-00146 Rome, Italy}\\[1mm]
{\it  (b) Dipartimento di  Fisica, Universit{\`a} di Roma  ``La Sapienza'' 
and \\  INFN, sezione di Roma, I-00185 Rome, Italy}\\[1mm]
 {\it (c) Physics Department, Brookhaven National Laboratory, \\Upton, New York
11973, US}\\[1mm]
\end{center}
\bigskip
\vspace{.5cm}

\centerline{\large\bf Abstract}
\begin{quote}
We present the constraints on  the trilinear Higgs self coupling
that arise from loop effects in  the $W$ boson mass  and the effective
sine predictions. We compute  
the contributions to these precision observables of  two-loop diagrams 
featuring an anomalous trilinear Higgs self coupling. We 
explicitly show that the same anomalous contributions are found if the
analysis of $\mw$ and $\sineff$ is performed  in a theory in which the 
scalar potential in the Standard Model Lagrangian is modified by an
(in)finite tower of $(\Phid \Phi)^n$ terms with $\Phi$ the Higgs doublet.  
We find that the bounds on the trilinear Higgs self coupling from precision 
observables are  competitive with those coming from Higgs pair production.
\end{quote}
\thispagestyle{empty}
\newpage

\section{Introduction\label{sec:intro}} 

%%%%%%%%%%%%%%%%
The discovery of a new scalar resonance with a mass around 125 GeV at
the Large Hadron Collider (LHC)\cite{Chatrchyan:2012xdj, Aad:2012tfa} 
opened a new era in high-energy particle physics. The study of the
properties of this particle provides strong evidence that it is the
Higgs boson of the Standard Model (SM), {\it i.e.}, a scalar CP-even
state whose coupling to the other known particles has a SM-like
structure and a strength proportional to their masses 
\cite{Khachatryan:2014jba,Aad:2015gba,Khachatryan:2016vau}.
At present, the  combined              
analysis based on 7 and 8 TeV LHC data sets \cite{Khachatryan:2016vau} shows
that the couplings with the vector bosons are found to be compatible with
those   expected from the SM within a              
$\sim 10 \%$ uncertainty, while in the case of the heaviest SM                  
fermions (the top, the bottom quarks and the $\tau$ lepton) the                 
compatibility is achieved with an uncertainty of $\sim 15-20              
\%$. Concerning the future, the best           
present estimates \cite{CMS:2013xfa, Peskin:2013xra} indicate that
at the end of the LHC            
Run-2 at $\sqrt{s}= 13-14$ TeV center-of-mass-energy,
the fit of the Higgs boson couplings to the vector bosons is              
expected to reach a $\sim 5 \%$ precision with 300 fb$^{-1}$                    
luminosity, while the corresponding ones for the fermions, with the             
exception of the $\mu$ lepton, can reach $\sim 10-15 \%$                        
precision. Similar estimates for the end of the High Luminosity option 
indicate a  reduction of these numbers by a factor $\sim 2$.

The study of the Higgs self interactions, coming from  the scalar potential
part in the Standard Model (SM) Lagrangian, is in a completely different 
status. In  the SM, the  Higgs potential in the unitary gauge reads
\beq
V(\phu) =  \frac{\mh^2}{2} \phu^2 + \tril v  \phu^3 +
\frac{\lambda_4}4 \phu^4  \label{eq:potun}
\eeq
where the Higgs mass ($\mh$) and the trilinear $(\tril)$ and quartic $(\qual)$ 
interactions are linked by the relations 
$\qual^{\rm SM}=\trilsm = \lambda =\mh^2/(2\,v^2)$,
where $v=(\sqrt2 \,G_\mu)^{-1/2}$ is the vacuum expectation value, and 
$\lambda$ is the coefficient of the $(\Phid \Phi)^2$
interaction, $\Phi$ being the Higgs doublet field.

The experimental verification of these relations, that fully characterize
the SM as a renormalizable Quantum Field Theory, relies on the  measurements 
of processes featuring at least two Higgs bosons in the final state. 
However, since the cross sections for this kind of processes are quite
small, constraining the Higgs self interaction couplings within few times their
predicted SM value is already extremely challenging.
In particular,  information on $\tril$ can be obtained from Higgs pair
production with  the present bounds on this reaction from 8 TeV data
that allow to constrain $\tril$ within ${\cal O}(\pm (15-20) \trilsm)$
\cite{Aad:2015xja,Aad:2015uka, Khachatryan:2016sey,ATLAS-CONF-2016-049}.
At $\sqrt{s} = 13 $ TeV, the Higgs pair production  cross section, in the SM,
is around 35 fb in the gluon-fusion
channel~\cite{Glover:1987nx,Plehn:1996wb,Dawson:1998py,Grigo:2013rya,
  Grigo:2014jma,Grigo:2015dia,Degrassi:2016vss,deFlorian:2013jea,
  Maltoni:2014eza,Borowka:2016ehy} and
even smaller in other production
mechanisms~\cite{Baglio:2012np,Frederix:2014hta}  that  suggests,
assuming an integrated luminosity of 3000 fb$^{-1}$,
that it will be possible to exclude at the LHC only values in the
range $\tril<-1.3 ~\trilsm$ and $\tril>8.7 ~ \trilsm$
via the $b\bar{b} \gamma \gamma$ signatures
\cite{ATL-PHYS-PUB-2014-019} or $\tril<-4 ~\trilsm$ and
$\tril>12 ~ \trilsm$  including also $b\bar{b} \tau
\bar{\tau}$ signatures \cite{ATL-PHYS-PUB-2015-046}.
Concerning  the quartic Higgs
self-coupling $\qual$,  its measurement via triple Higgs production
seems beyond the reach  of the LHC\cite{Plehn:2005nk, Binoth:2006ym} due to
the smallness of the corresponding cross
section (around $0.1$ fb) \cite{Maltoni:2014eza}.

In order to  constrain the trilinear Higgs self coupling,
a complementary strategy based  on the precise
measurements of single Higgs production and decay processes was recently
proposed. In this approach the effects induced at the loop level on
single Higgs processes by a modified $\tril$ coupling are studied.
This approach builds on the assumption that New Physics (NP) couples to
the SM via the Higgs potential in such a way that the lowest-order Higgs
couplings to the other fields of the SM (and in particular to the top quark and
vector bosons) are still given by the SM prescriptions or,
equivalently, modifications to these couplings are so small that do
not swamp the loop effects one is considering.
This strategy was first applied to $ZH$ production at an $e^+ e^-$ collider in
Ref.~\cite{McCullough:2013rea} and later to Higgs production and decay modes
at the LHC \cite{Degrassi:2016wml,Gorbahn:2016uoy,Bizon:2016wgr}. 

The aim of this  work is twofold. On the one side we apply the same strategy
to the precise measurements 
of the $W$ boson mass, $\mw$,  and the effective sine, $\sineff$. 
In order to constrain $\tril$ we
look for effects induced by an anomalous Higgs trilinear coupling at
the loop level in the predictions of $\mw$ and $\sineff$. Following the 
approach of
Ref.~\cite{Degrassi:2016wml} we parametrize the effect of NP at the weak scale
via a single parameter $\ktre$, {\it i.e.}
the rescaling of the SM trilinear coupling $\trilsm$, so that the $\phu^3$
interaction in the  potential  is given by
\beq
V_{\phu^{3}} = \tril\, v \,\phu^3 \equiv \ktre \trilsm \, v \, \phu^3\,,
~~~~~~~\trilsm\equiv\frac{G_\mu}{\sqrt{2}} \mh^2 \,,
\label{h3coeff}
\eeq
and compute, in the unitary gauge, the effects induced by $\ktre$ in  the
two-loop $W$ and $Z$ boson self-energies, which are the relevant quantities
entering in the two-loop determination of $\mw$ and $\sineff$. On the
other side we specify better the anomalous coupling approach employed
above by showing that, at the order we are working, {\it i.e.}
at the two-loop level, it is equivalent to the use of a  SM
Lagrangian with a  scalar potential given by an (in)finite tower
of $(\Phid \Phi)^n$ terms. 
Furthermore, we show that the use of the unitary
gauge in the anomalous coupling approach does not introduce any gauge-dependent
problematics.

The paper is organised as follows. In Section \ref{sec:2} we discuss the
contributions induced by an anomalous Higgs trilinear coupling in $\mw$
and $\sineff$. Section \ref{sec:3} is devoted to show that the addition
to the SM Lagrangian  of $(\Phid \Phi)^n$ terms gives rise to the
same contributions. In the following section we discuss the constraints
on $\lambda_3$ that 
can be obtained from the current data. In the last section we summarise and 
draw our conclusions.  

\section{$\tril$-dependent  contributions in $\mw$ and $\sineff$}
\label{sec:2}
We consider a Beyond-the-Standard-Model (BSM) scenario, described at 
low energy by the SM Lagrangian with a modified scalar potential. 
We further assume that only Higgs self couplings will be affected
by this modified potential while the strength of the couplings of the Higgs to 
fermions and vector bosons will not change with respect to its SM value, or,
equivalently, that any modification of these couplings is going to induce
effects much smaller than the ones coming from the ``deformation'' of the
Higgs self couplings.

In the \bms\ formulation of the radiative corrections
\cite{Sirlin:1989uf,Fanchiotti:1989wv,Degrassi:1990tu} the theoretical
predictions of $\mw$ and $\sineff$ are expressed   in terms of the pole
mass of the particles, the \bms\
Weinberg angle $\hat{\theta}_{W} (\mu)$ and the
\bms\ electromagnetic coupling $\alc (\mu)$, defined at the 't-Hooft
mass scale $\mu$, usually chosen to be equal to $\mz$. In particular,
given  the radiative parameters  $\drcarw$, $\Delta \alc$, $Y_{\ms}$ 
defined through  ($\sincur (\mz)\equiv \scs$) \cite{Degrassi:2014sxa}
\beqn
&& \frac{G_\mu}{\sqrt{2}} = \frac{\pi \alc
  (\mz)}{2 \mw^2 \scs} \left( 1 + \drcarw \right), \label{drcarw}
\qquad \alc (\mz)= \frac{\alpha}{1-\Delta \alc (\mz)}, \nonumber\\ &&
\quad  \qquad\rhoh \equiv \frac{\mw^2}{\mz^2 \ccs} =\frac{1}{ 1 - Y_{\ms} }~,
\label{rhoc}
\eeqn
with $\ccs = 1- \scs$, $\mw $ is obtained from $\mz, \alpha, G_\mu$ via
\beq
        \mw^2 = \frac{\rhoh\, \mz^2}{2} \left\{
         1 + \left[ 1 - \frac{4 \hat{A}^2}{\mz^2 \rhoh} (1+\drcarw) \right]^{1/2}
                            \right\}~ ,
\label{mwcur}
\eeq
where $\hat{A}= (\pi \alc (\mz)/(\sqrt{2} G_\mu))^{1/2}$,
while the effective sine is related to $\scs$ via
\beq
\sineff =  \kcar_\ell(m_Z^2) \scs,~~~~
\kcar_\ell(\mz^2)= 1 + \delta \kcar_\ell(\mz^2) ,
\label{sineff}
\eeq
where $\kcar_\ell(q^2)$ is an electroweak form factor\footnote{In our \bms\
formulation the top contribution is not decoupled. Then $\kcar$ is very close
to 1 and $\sineff$ can be safely identified with $\scs$ ~\cite{Gambino:1993dd}.}
(see Ref.~\cite{Gambino:1993dd}) and
\beq
        \scs = \frac{1}{2} \left\{
         1 - \left[ 1 - \frac{4 \hat{A}^2}{\mz^2 \rhoh} (1+\drcarw) \right]^{1/2}
                            \right\} ~.
\label{sincur} 
\eeq

In our BSM  scenario the modifications of the scalar potential affect the 
radiative parameters $\drcarw$ and $Y_{\ms}$ at the two-loop level while 
$\Delta \alc$ and $\delta \kcar_\ell(\mz^2)$ are going to be affected only at 
three loops.
Recalling that the present knowledge of $\mw$ and $\sineff$ in the SM includes
the complete two-loop corrections, we are going to discuss only the
modifications induced  in $\drcarw$ and $Y_{\ms}$. The two-loop contribution
to $\drcarw$ and $Y_{\ms}$ can be expressed as \cite{Degrassi:2014sxa}
\beqn
\drcarwd &=&
\frac{{\rm Re} \,\aww^{(2)}( \mw^2)}{\mw^2} -\frac{\aww^{(2)}(0)}{\mw^2} + \dots
\label{drcarw2}\\
 Y^{(2)}_{\ms} & = &  {\rm Re} \left[
   \frac{\aww^{(2)}(\mw^2)}{\mw^2} -  \frac{\azz^{(2)}(\mz^2)}{\mz^2}
    \right] +   \dots
 \label{Yms2l}
 \eeqn
 where $\aww \: (\azz)$ is the term proportional to the metric tensor in the
 $W\, (Z)$ self energy with  the superscript indicating the loop order,
 and the dots represent additional two-loop
 contributions that are not sensitive to a modification of the scalar
 potential.

From the knowledge of the additional contributions
induced in $\drcarwd$ and $ Y^{(2)}_{\ms} $ one can easily obtain the modification
of the radiative parameters $\Delta r$ and $\kappa_e (\mz^2)$ of the On-Shell
(OS) scheme \cite{Sirlin:1980nh}. Considering only new contributions
from the modified scalar potential one can write
\beq
\Delta r^{(2)}  = \drcarwd -\frac{c^2}{s^2} Y^{(2)}_{\ms} \, ,
\label{deltar}
\eeq
where $c^2 \equiv \mw^2/\mz^2$, $ s^2 = 1 - c^2$ with $\Delta r$ being the
radiative parameter entering the $\mw-\mz$ interdependence. The effective
sine is related to $s^2$ in the OS scheme via $\sineff = \kappa_e (\mz^2) s^2$
and for the new contributions in  $\kappa_e (\mz^2)$ one can  write
\beq
\kappa^{(2)}_e (\mz^2) = 1 -\frac{c^2}{s^2} Y^{(2)}_{\ms}~.
\label{kappos}
\eeq
 
\begin{figure}[t]
\begin{center}\vspace*{-1.0cm}
\includegraphics[width=1.0\textwidth]{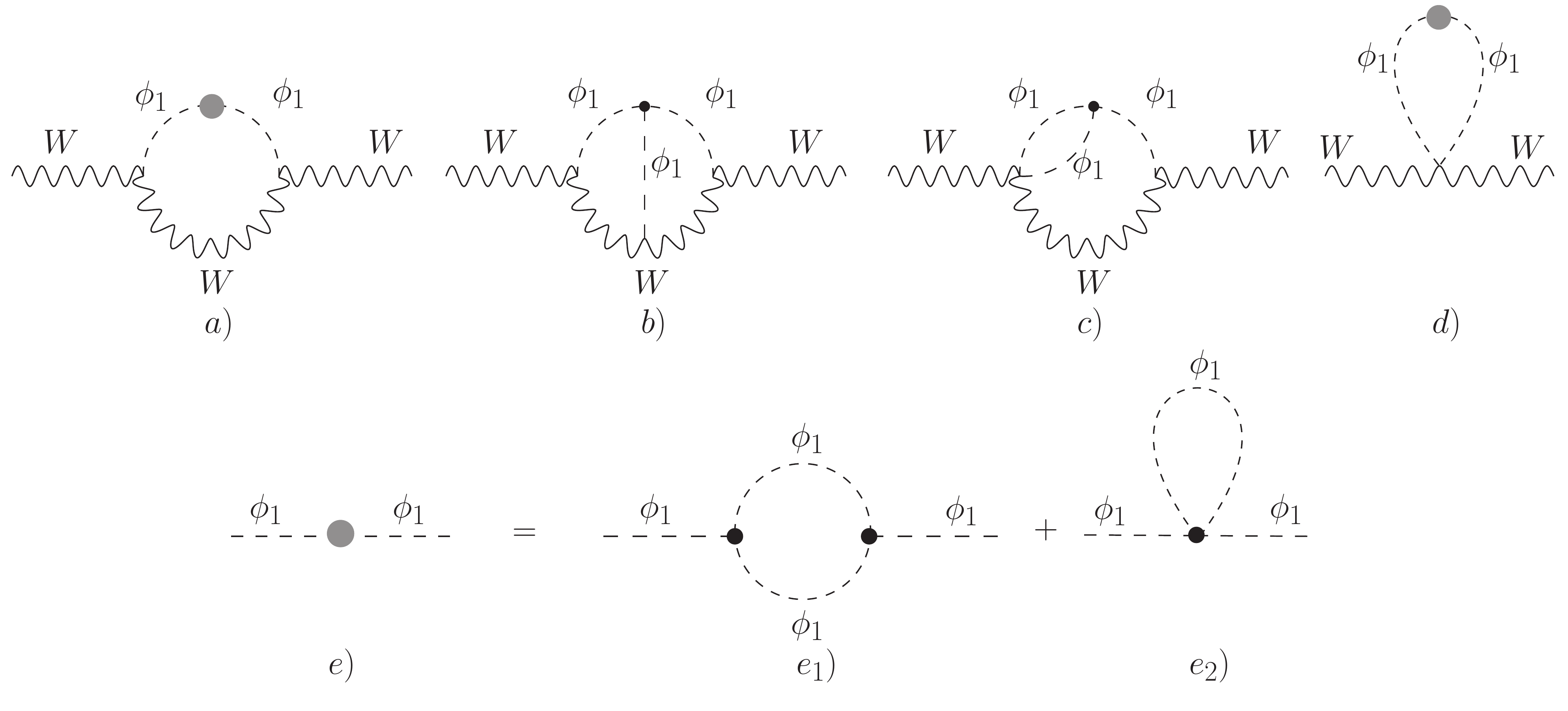}
\ \vspace*{-1.0cm}
\caption{ Two-loop $\tril$-and-$\lambda_4$-dependent diagrams in  the $W$ 
self-energy, in the unitary gauge. The dark blob represent the insertion
of the modified diagrams in the one-loop Higgs self energy, shown in the
second row. The black point represents either an  anomalous
$\tril$ or $\qual$.}
\label{fig:1}
\end{center}
\end{figure}

The new contribution  in the self energies in eqs.~(\ref{drcarw2},\ref{Yms2l})
can be parametrized just by a 
modification of  the trilinear coupling as described in eq.~(\ref{h3coeff}).
In order to correctly identify the effects related to the $\phu^3$ 
interaction we follow Ref.~\cite{Degrassi:2016wml} and work in the unitary
gauge. Here we discuss the $W$ self energy but an identical analysis can be done
also for the $Z$ self energy. 

The two-loop diagrams in the $W$ self energy
that are sensitive to a modification of the Higgs self couplings are depicted
in fig.~\ref{fig:1}. The dark blob in diagrams \ref{fig:1}$a)$, \ref{fig:1}$d)$
represents the one-loop Higgs self energy or the one-loop Higgs mass
counterterm that in our scenario gets
modified with respect to the SM result in the unitary gauge by the diagrams in 
fig.~\ref{fig:1}$e)$. The  amplitudes of the diagrams in fig.~\ref{fig:1}
were generated using the Mathematica package {\sc\small FeynArts}
\cite{Hahn:2000kx} and  reduced to  scalar Master Integrals using private
codes and the packages FeynCalc \cite{Mertig:1990an,Shtabovenko:2016sxi } and
Tarcer \cite {Mertig:1998vk}. After the reduction to scalar integrals we were
left with the evaluation of two-loop vacuum integrals and
two-loop self-energy diagrams at external momenta different from zero.
The former integrals were evaluated analytically using the results
of Ref.\,\cite{Davydychev:1992mt}. The latter ones  were instead
reduced to the set of loop-integral basis functions introduced
in Ref.\,\cite{Martin:2003qz}. For their numerical evaluation we used the C
program TSIL \cite{Martin:2005qm}. Our results are expressed in terms of
the OS Higgs mass that specifies the Higgs mass counterterm.

Few observations are in order: i) the insertion of the ``cactus''
diagram $e_2)$ in diagrams $a)$ and $d)$ in fig.~{\ref{fig:1} gives
  rise to a contribution proportional to the quartic Higgs self
  couplings on which we did not make any assumption. However, this
  contribution is exactly cancelled by the corresponding Higgs mass
  counterterms diagram so that the final result does not depend on
  $\lambda_4$.  This finding is general and does not depend on the
  particular scheme used to define the Higgs mass.  Using a
  different Higgs mass definition, like, e.g., an $\bms$
  Higgs mass, $\hat{m}_{\sss H}$, the expression for the $W$ self-energy
  acquires an explicit $\lambda_4$ dependence. However, this
  dependence is going to be cancelled by the $\lambda_4$ dependence of
  $\hat{m}_{\sss H}$, when the latter is extracted from a physical quantity
  like the OS mass.
ii) We expect the modified potential to contain 
Higgs self  interactions with a number of $\phu$ fields larger than 4 (quintic,
 sextic, etc. interactions). However, none of these interactions is going to 
 contribute to the $W$ self energy at the two-loop level\footnote{A quintic
self interaction gives rise to a two-loop tadpole. However, tadpole 
contributions cancel in eqs.~(\ref{drcarw2},\ref{Yms2l}).}. Thus the new
 contributions induced by our BSM scalar potential at the two-loop level
 are only functions of $\ktre$. iii) The contribution to the physical 
observables given by the diagram \ref{fig:1}$d)$   vanishes 
in the differences of self energies (see eqs.~(\ref{drcarw2},\ref{Yms2l})).

As in the case  of single Higgs processes the $\tril$-dependent contributions
can be divided into a part quadratically dependent on $\tril$ and another
linearly proportional to $\tril$. The former is due to the diagram 
\ref{fig:1}$a)$ with the insertion of diagram \ref{fig:1}$e_1)$ and
of its corresponding  Higgs mass counterterm.
The latter is given by diagrams \ref{fig:1}$b)$, \ref{fig:1}$c)$.

\section{Equivalence with a $(\Phid \Phi)^n$ theory}
\label{sec:3}
In this section we show that the results presented in section \ref{sec:2},
where no specific assumption on  the BSM scalar potential
was made, can be obtained using a SM Lagrangian with a scalar
potential of the form
\beq
V^{NP} = \sum_{n=1}^N c_{2n} ( \Phid \Phi )^n\,,  \qquad\qquad 
\Phi = \binom{\php}{\frac{1}{\sqrt{2}}( v+ \phu + i \phd )}\,,
\label{potential}
\eeq
where $N$ can be a finite integer or infinite, and in the latter case we
assume the series to be convergent. This is the only constraint we impose
on the $c_{2n}$ coefficients, in particular we do not assume an
effective-field-theory (EFT) scaling on them, 
i.e.~$c_{2n+2} \sim c_{2n}/\Lambda^2 $
with $\Lambda$ the scale of NP.
The SM potential is recovered setting $N=2$ in eq.~(\ref{potential}) 
with $c_2 = - m^2$ and  $c_4 = \lambda$ where $-m^2$ is the Higgs mass term in
the SM Lagrangian in the unbroken  phase. 

Defining $ \phi_{2u} = \php\phm + \frac12 \phd^2$ the $n$-th term in the series 
can  be  written as
\beq
( \Phid \Phi )^n =  \sum_{k=0}^n \sum_{j=0}^k \sum_{h=0}^j \binom{n}{k} 
\binom{k}{j} \binom{j}{h} \phi_{2u}^{n-k}\left(\frac{v^2}{2}\right)^{k-j}
\left(\frac{\phu^2}{2}\right)^{j-h}(v\phu)^h ~,
 \label{nterm}
\eeq
with
\beq
 \binom{n}{k} 
\binom{k}{j} \binom{j}{h} =  \frac{n!}{(n-k)!(k-j)!(j-h)! h!}~,
\eeq
and its contribution to any Higgs self interaction can be labelled by
the triplet $\{k,j,h\}$. For example, the minimum of the potential can be 
obtained from the triplet $\{n,1,1\}$:
\beq
\left.\frac{d \, V^{NP}}{d \, \phu} \right|_{\phu=0}=
v\,\sum_{n=1}^N c_{2n}\,  n \left(\frac{v^2}{2}\right)^{n-1} =0 ~,
\label{minV}
\eeq
while  the Higgs mass is given by the two triplets $\{n,1,0\}$ and $\{n,2,2\}$.
However, due to the condition in eq.~(\ref{minV}), the
first one is giving a vanishing contribution so that 
\beq
\mh^2 = 
v^2\,\sum_{n=1}^N c_{2 n}\,  n(n-1) \left(\frac{v^2}{2}\right)^{n-2} ~.
\label{Hmass}
\eeq
The potential $V^{NP}$ up to quartic interactions can be written as
%\beqn
%V_{4\phi}^{NP} &=& \frac{\mh^2}{2v^2}\,\xi^2 + \left( \frac{\mh^2}{2 v^2} + 
%d \lambda_4\right)
%\frac14\,\phu^4 + \left( \frac{\mh^2}{2 v^2} + 3 \,d \tril \right)\,\xi\,\phu^2 \nonumber\\
%&& + 
%\left( \frac{\mh^2}{2 v}  + d \tril\right)\phu^3 +\frac{\mh^2}{v}\,\xi\,\phu 
 % +\frac12 \mh^2 \,\phu^2\,,
%\label{PotqP}
%\eeqn
\beqn
V_{4\phi}^{NP} &=& \frac{\mh^2}{2v^2} \left [ \php \phm (\php \phm + \phd^2) +
 \frac14 \phd^4 \right ] + \left( \frac{\mh^2}{2 v^2} + d \lambda_4\right)
\frac14 \phu^4 \nonumber\\
&& + \left( \frac{\mh^2}{2 v^2}  + 3 \,d \tril \right)  \phu^2 
\left[ \php \phm  + \frac12 \phd^2 \right ] + 
\left( \frac{\mh^2}{2 v}  + v\,d \tril\right)\phu^3 \nonumber\\
 && +\frac{\mh^2}{2 v} \,\phu \left (\phd^2 + 2 \php\phm \right ) 
 +\frac12 \mh^2 \,\phu^2\,.
\label{PotqP}
\eeqn
with
\beqn
d \tril &=& 
\frac13 \sum_{n=3}^N c_{2n} \, n(n-1)(n-2) \left(\frac{v^2}{2}\right)^{n-2} \,,
\label{dc3}\\
d \lambda_4 & =& \frac23 \sum_{n=3}^N c_{2n} \, n^2(n-1)(n-2)  
\left(\frac{v^2}{2}\right)^{n-2} ~.
\label{dc4}
\eeqn
It is worth  noting that in eq.~(\ref{PotqP}) only few couplings are
modified with respect to their SM values. In particular, concerning the 
unphysical scalars, only the coupling of $\phi_{2u}\,\phu^2$ is modified, with a 
deformation that is related to the deformation of $\tril$.
  
\begin{figure}[t]
\begin{center}%\vspace*{-1.0cm}
\includegraphics[width=1.0\textwidth]{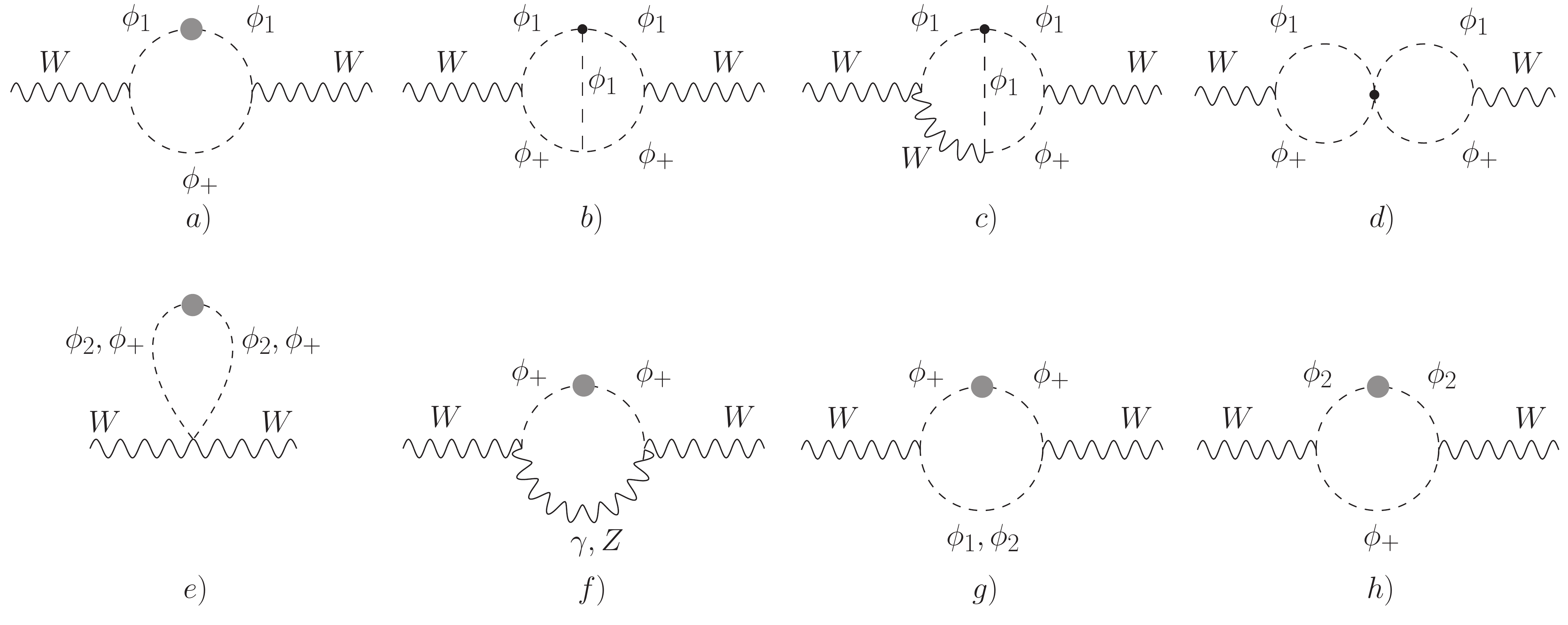}
\caption{ Two-loop diagrams in   the $W$  self-energy, involving unphysical 
scalars where modified couplings (black points) from $V_{4\phi}^{NP}$ appear. The 
dark blob  represents  the insertion of the relevant one-loop self energy 
(see fig.~\ref{fig:2}).}
\label{fig:3}
\end{center}
\end{figure}

\begin{figure}[t]
\begin{center}%\vspace*{-1.0cm}
\includegraphics[width=1.0\textwidth]{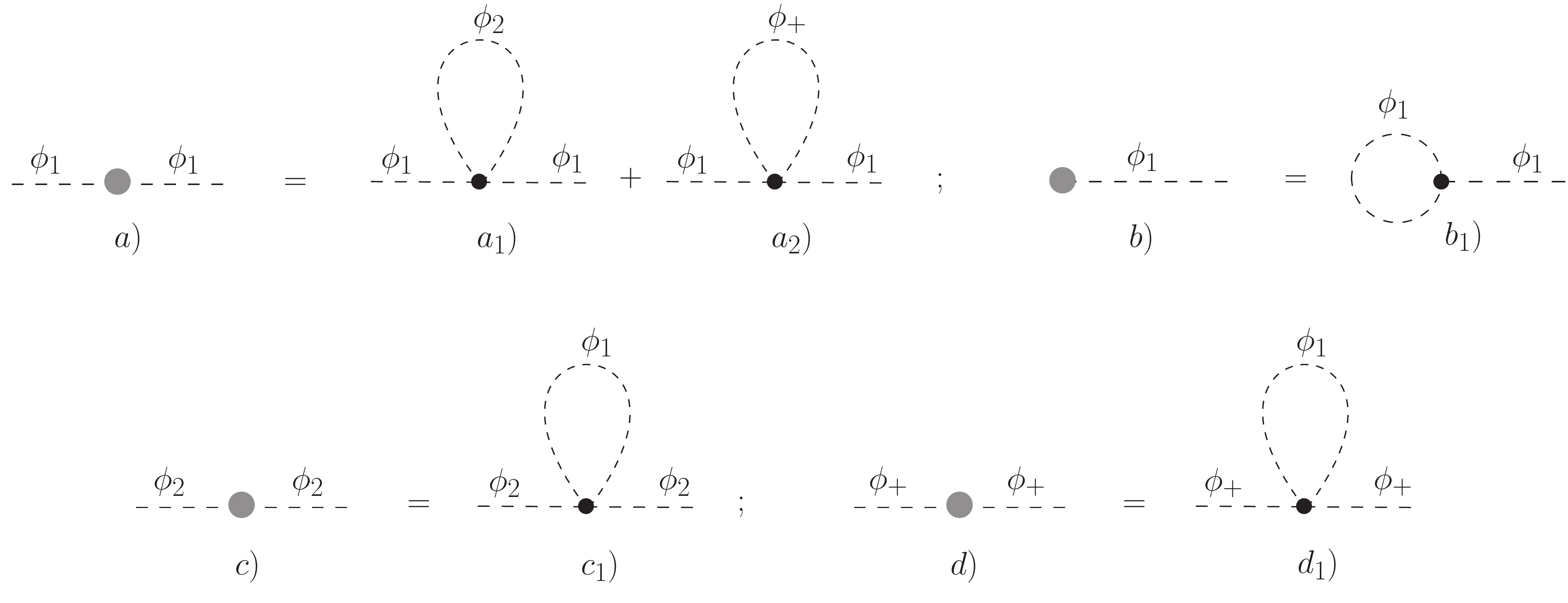}
\ \vspace*{-1.0cm}
\caption{ One-loop self energy and tadpole diagrams that contain modified
couplings with respect to the SM.}
\label{fig:2}
\end{center}
\end{figure}

In order to show that the result for the two-loop $W$ self energy computed 
using $V^{NP}$ is egual to the one obtained assuming an anomalous $\tril$ 
working in the unitary gauge, we have to analyze the two-loop diagrams
that are modified with respect to their SM result working in a generic
$R_\xi$ gauge.  Besides the ones in fig.~\ref{fig:1}, now computed in an
$R_\xi$ gauge, the   diagrams containing  unphysical scalars, shown in
fig.~\ref{fig:3},  should be taken into account. In the latter figure  the dark
blob represents the insertion of the relevant one-loop self energy. 
In fig.~\ref{fig:2} we show for the various self energies and the tadpole only
the diagrams that are modified with respect to their SM result due to
the new scalar potential $V^{NP}$. It easy to show that the only 
non-vanishing contributions in figs.~\ref{fig:1}$a)$, \ref{fig:1}$d)$,
\ref{fig:3}$a)$  come from  the insertion of diagram $e_1)$ in
fig.~\ref{fig:1} plus its corresponding counterterm diagram while all
the other insertions being of the cactus type (see \ref{fig:1}$e_2)$ and  
\ref{fig:2}$a)$) are cancelled against the corresponding Higgs mass counterterm
diagrams. 
Furthermore the sum of  diagrams \ref{fig:1}$a)$ and  \ref{fig:3}$a)$
is gauge invariant. Similarly one can prove that the sum of diagrams
\ref{fig:1}$b)$, \ref{fig:1}$c)$, \ref{fig:3}$b)$, and \ref{fig:3}$c)$, is gauge
invariant.

To complete our proof about the equivalence of the two computations we
have to show that the additional 
contributions with respect to the SM results in the diagrams
\ref{fig:3}$d)$--\ref{fig:3}$h)$
and in  the corresponding counterterm diagrams must vanish.   
Diagram \ref{fig:3}$d)$ is automatically zero, while in the remaining diagrams 
a self energy of an unphysical scalar is always present. According to
$V_{4\phi}^{NP}$  the only modified contributions in the one-loop self energies
of the unphysical scalars are given by diagrams \ref{fig:2}$c_1)$ and
\ref{fig:2}$d_1)$. To the contribution of diagrams
\ref{fig:3}$e)$--\ref{fig:3}$h)$ with
the insertion of \ref{fig:2}$c_1)$ or \ref{fig:2}$d_1)$ one has to add
the counterterm diagrams.  The counterterm associated to the renormalization of
the mass of an unphysical scalar contains a term related to the mass of
the corresponding vector boson plus a term that is related to the 
renormalization of the vacuum. The former is not affected by our modified
scalar potential. The latter, when  $v$ is identified with the 
minimum of the radiatively corrected  potential,  is given by the tadpole
contribution \cite{Sirlin:1985ux}. Then the only modified contribution
in the mass renormalization of the unphysical scalars is given by diagram
\ref{fig:2}$b_1)$. Thus,  the additional  contributions with respect to
the SM result in  the diagrams \ref{fig:3}$e)$-\ref{fig:3}$h)$ are exactly 
cancelled by
the additional contributions in the unphysical scalar mass counterterm diagrams.
The key point in this cancellation is the fact that  the modification
in the  vertex with three physical Higgses and the one in the vertices
containing  two physical and two unphysical Higgses are related by a
factor $3/v$ as shown in eq.~(\ref{PotqP}).

We have shown that  in a theory with a scalar potential given by 
eq.~(\ref{potential}) the  two-loop $W$ self energy is modified with
respect to its SM value by additional contributions that are gauge-invariant.
Then, one can directly compute them in the unitary gauge, that corresponds to
the computation with an anomalous $\tril$ once the identification
$\ktre = 1 + 2 v^2/\mh^2 \, d\tril$ is made.

\section{Results}
\label{sec:4}
The analytic expressions for the contributions induced in $\drcarwd$
and $Y^{(2)}_{\ms}$ by an anomalous $\tril$ are reported in the Appendix.
These contributions are going to modify the SM predictions for  $\mw$
and $\sineff$ via eqs.~(\ref{drcarw}--\ref{sincur}). 

Denoting as $O$ either $\mw$ or $\sineff$ one can write
\beq
O  =  O^{\rm SM} \left[ 1 + (\ktre-1) C_1+(\ktre^2-1) C_2 \right] \, , 
\label{corr}
\eeq
with  the values of the coefficients $C_1$ and $C_2$  reported in 
Table \ref{tab:1}. 

\begin{table}[t]%[ph]
\begin{center}
\begin{tabular}{c|r|r|}
      & $C_1$~~~~~ & $C_2$~~~~~ \\ 
\hline
$\mw$ & $ 6.27\times 10^{-6}$  & $-1.72 \times 10^{-6}$ \\
$\sineff$ & $ -1.56 \times 10^{-5}$ & $4.55  \times 10^{-6}$
\end{tabular}
\end{center}
\caption{ Values of the coefficients $C_1$ and $C_2$.}
\label{tab:1}
\end{table}

Let us comment on the validity of eq.~(\ref{corr}). At the two-loop level
we are working, the contributions 
induced by an anomalous Higgs trilinear coupling  in the  precision observables
are finite (see table \ref{tab:1} or  the Appendix), i.e.  they are not 
sensitive to the NP scale $\Lambda$ associated with the modification of the 
potential.
This  situation is analogous  to  what happens in single Higgs processes
where new contributions induced by an anomalous $\tril$ at the NLO
are also finite \cite{Degrassi:2016wml}. As in single Higgs processes if
NNLO effects are considered, one expects that at  three or more loops    
the modified potential is going to induce contributions not only proportional
to $\tril$ but also to quartic, quintic etc. Higgs self interactions and
moreover these contributions will be sensitive to the NP scale. 

The constraints on $\ktre$ we are going to derive below assume the validity
of a perturbative approach. Then, we expect  any higher-order contribution
to be subdominant with respect to the effects we are computing.
This implies that these higher-order contributions should not contain any
large amplifying factor related to the scale $\Lambda$, or 
equivalently that  $\Lambda$ cannot be too far 
from the Electroweak scale. Furthermore, since at the three-loop level
one expects the  anomalous contribution from the trilinear coupling to grow 
as $\ktre^4$, a restricted range of $\ktre$ should also be imposed. Following
Ref.~\cite{Degrassi:2016wml}  we consider $|\ktre| \lesssim 20$ as a range of
validity of our perturbative approach. 

In order to set limits on $\ktre$ from the analysis of precision observables, 
we perform a simplified fit. 
We define the best value of $\ktre$ as the one that minimizes the 
$\chi^2(\ktre)$ function defined as
\begin{equation}\label{chi2}
\chi^2(\ktre)\equiv 
 \sum \frac{(O_{\rm exp}-O_{\rm the})^2}{(\delta)^2}\,,  
\end{equation}
where $O_{\rm exp}$ refers to the experimental measurement of the observable 
$O$, $O_{\rm the}$ is its theoretical value obtained from eq.~(\ref{corr}) and 
$\delta$ is the total uncertainty,  that we take as the sum in quadrature
of the experimental and theory errors. In order to ascertain the goodness of 
our fit, we also compute the $p$-value as a function of $\ktre$:
\beqn 
\text{$p$-value}(\ktre)=1-F_{\chi^2_{(n)}}(\chi^2(\ktre))\,,
\eeqn
where $F_{\chi^2_{(n)}}(\chi^2(\ktre))$ is the cumulative distribution function 
for a $\chi^2$ distribution with $n$ degrees of freedom, computed at 
$\chi^2(\ktre)$.

\begin{figure}[t]
\begin{center}%\vspace*{-1.0cm}
\includegraphics[width=0.45\textwidth]{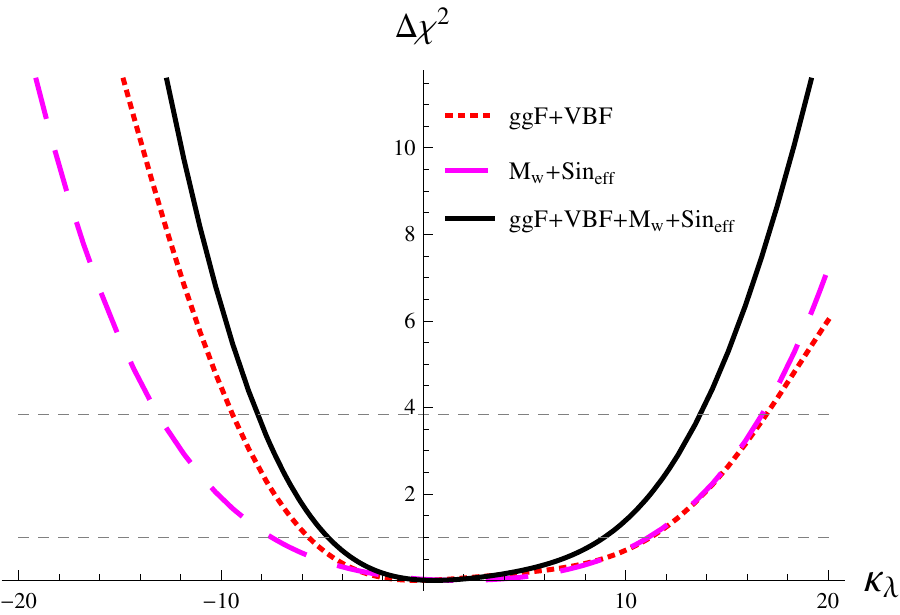}
\includegraphics[width=0.45\textwidth]{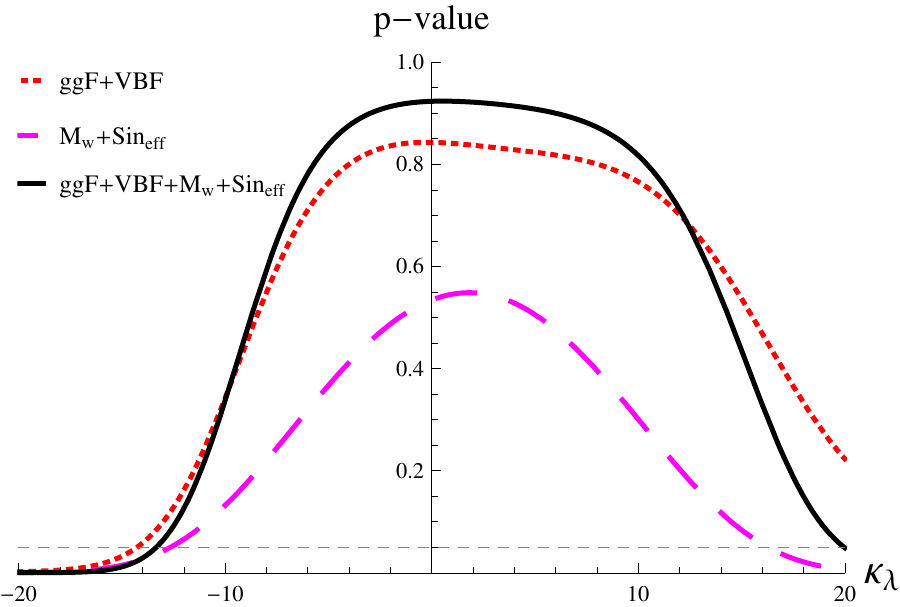}
\caption{Left: $\chi^2$ for the different sets of observables described in the 
text, the two horizontal lines represent $\Delta \chi^2=1$ and 
$\Delta \chi^2=3.84$. Right: corresponding $p$-value, the horizontal line 
is $p=0.05$.}
\label{fig:4}
\end{center}
\end{figure}

In the fit we consider not only the two precision observables but also the
signal strength parameter for single Higgs production in gluon fusion ($\ggF$)
and vector boson fusion (VBF). The latter observables were indicated as the 
$P_2$ set in Ref.~\cite{Degrassi:2016wml} where it was shown that they were
returning the most stringent bound on $\ktre$.   
We then considered three set of data: 
\begin{itemize}
\item 
The $P_2$ set in Ref.~\cite{Degrassi:2016wml}. The experimental results are 
presented in Tab.~8 of Ref.~\cite{Khachatryan:2016vau}. See 
Ref.~\cite{Degrassi:2016wml} for more details. 
\item 
The W mass and effective sine. For the W mass we use the latest result 
by the ATLAS collaboration
$\mw = 80.370 \pm 0.019$ GeV \cite{Aaboud:2017svj}. This number, although
it has a slightly larger uncertainty with respect to the world average
$\mw = 80.385 \pm 0.015$ GeV \cite{Olive:2016xmw}, it is closer to the SM 
prediction $\mw = 80.357 \pm 0.009 \pm 0.003$ where the errors refer to the
parametric and theoretical uncertainties \cite{Degrassi:2014sxa}. Concerning
the effective sine, we use the average of the CDF \cite{Aaltonen:2016nuy}
and D0 \cite{Abazov:2014jti} combinations
$\sineff = 0.23185 \pm 0.00035$ \cite{Olive:2016xmw}, to confront against the 
SM result $\sineff=0.23145\pm 0.00012 \pm 0.00005$, where again the errors 
refer to parametric and theoretical uncertainties 
respectively \cite{Degrassi:2014sxa,Awramik:2006ar}.
\item 
The combination of these two sets of data.
\end{itemize}

The $\chi^2(\ktre)$ and $p$-value functions for the three sets are reported in 
fig.~\ref{fig:4}. In particular for the combination we find
\beqn \ktre^{\rm best}=0.5\,, ~~~~~~
\ktre^{1 \sigma} = [-4.7,8.9]\,, ~~~~~~ \ktre^{2 \sigma} =
     [-8.2,13.7]\,,
\label{Knum}
\eeqn 
where the $\ktre^{\rm best}$ is the best value and $\ktre^{1 \sigma}$, 
$\ktre^{2 \sigma}$ are respectively the $1\sigma$ and
$2\sigma$ intervals. We identified the $1\sigma$ and
$2\sigma$ intervals assuming a $\chi^2$ distribution.
The comparison between the numbers in eq.~(\ref{Knum}) and the corresponding
ones for the $\ggF+$VBF case \cite{Degrassi:2016wml}, namely
\beqn \ktre^{\rm best}=-0.24\,, ~~~~~~
\ktre^{1 \sigma} = [-5.6,11.2]\,, ~~~~~~ \ktre^{2 \sigma} =
     [-9.4,17.0]\,~~~ (P_2\: {\rm set}),
\label{KmunP2}
\eeqn 
 shows that  the inclusion of the precision observables reduces the
allowed range for $\ktre$.
Similarly, looking at the solid black line in the $p$-value part of 
fig.~\ref{fig:4}, we can exclude at more than $2\sigma$ models with
$\ktre$ in the regions  $\ktre<-13.3$ and $\ktre>20.0$.
 
 These results indicate  that in the future, when  more accurate
measurements  will be available,
the combination on $\mw$ and $\sineff$ with  single Higgs processes
could be very powerful in constraining  the allowed region for $\ktre$, in 
particular  the region of positive $\ktre$.

\section{Conclusions}
In this work we have discussed how the predictions of the $W$ boson mass and
the effective sine are affected by  loops featuring  an anomalous
trilinear Higgs coupling. Following Ref.\cite{Degrassi:2016wml} we
have chosen to present our results in the contest of the
$\kappa$-framework,  parametrising  the effect of NP at the weak scale via a 
single parameter, $\ktre$, i.e. the rescaling of the SM Higgs trilinear 
coupling. Indeed, 
given a generic scalar potential constructed using only the Higgs doublet 
field, at the two-loop level these precision observables are only sensitive to 
the modification of the trilinear Higgs coupling. 
As in Ref.\cite{Degrassi:2016wml}
we worked in the unitary gauge to easily identify the effects we were
looking for. We  proved that the latter choice is just a technical
trick and does not introduce any gauge-dependent issues. In fact, 
we have explicitly shown that our approach is
equivalent, to the order we were working,
to an analysis of $\mw$ and $\sineff$ in
a generic $R_\xi$ gauge  performed  in a theory described
by a SM Lagrangian in which  the scalar potential is modified by the
addition of an (in)finite tower of $(\Phid \Phi)^n$ terms. 

Concerning this scalar potential, 
one important point to remark is the fact that  we did not make any assumption 
on the size of the coefficients of the various terms in the potential, 
so that in principle we do not have a priori any restriction on $\ktre$, apart from the requirement of perturbativity. This
is at variance with an EFT approach based on the addition to the SM Lagrangian
of a dimension six $(\Phid \Phi)^3$ term \cite{Gorbahn:2016uoy,Bizon:2016wgr}
where the requirement of $v$ being a global minimum constraints $\ktre < 3$
\cite{Grojean:2004xa,Huang:2015tdv}. 

However, in order to keep under control higher-order effects
induced by quartic, quintic etc. Higgs self interaction  some relations
among the $c_{2n}$ coefficients in the potential should be assumed. 
In particular, either one assumes
that  $c_{2n}$ exhibit a scaling with the order $n$ so that
the couplings of the interactions with a large number of $\phu$ do not
grow (as an  example $d \qual$ does not become larger than $d \tril$, see 
eqs.~(\ref{dc3},\ref{dc4})) or that the various $c_{2n}$
are related to each other enforcing cancellations among the various terms in the
potential.

 As we said a theory with a modified trilinear coupling is expected to
 be valid up to a scale $\Lambda$ that cannot be too far from the
 Electroweak scale. An estimate of $\Lambda$ can be obtained by
 looking when perturbative unitarity is lost in processes like
 e.g. the annihilation of longitudinal vector bosons into $n$ Higgs
 bosons, $V_L V_L \to n \, \phi_1$ \cite{Cornwall:1974km}. A preliminary study on this subject indicates
  that $\Lambda \sim 1-3$ TeV  \cite{Falkowski}.

We have estimated the sensitivity of $\mw$ and $\sineff$ to an anomalous
trilinear coupling via a one-parameter fit. We have also shown that when the
 analysis of the precision observables is combined with the one from single
Higgs inclusive measurements at the LHC 8 TeV, a restricted range of allowed
$\ktre$ is found. The range found is actually competitive with the
present bounds obtained from the direct searches of Higgs pair production.

\section*{Acknowledgements}
Two of us (G.D., P.P.G.) want to thank their  collaborators F.~Maltoni
and D.~Pagani together with Xiaoran Zhao for useful discussions and 
communications. G.D.~wants to thank R.~Rattazzi for an enlightening ``endless''
discussion. The work of P.P.G. is supported by the United States Department of 
Energy under Grant Contracts de-sc0012704.

\clearpage
\appendix
\begin{appendletterA}

  \section*{Anomalous contributions in $\drcarwd$ and $Y^{(2)}_{\ms}$}
  Here we give the analytic expressions for the additional contributions induced
  in $\Delta \hat{r}^{(2)}_{W}$
and $Y^{(2)}_{\ms}$ by an anomalous $\tril$. In the formulae below
\beq 
\mhw=\frac{\mh^2}{\mw^2}\,,  \quad \lnb(x)=\log\left(\frac{x}{\mu}\right)\,,
\eeq
with $\mu$ the 't-Hooft mass scale. We find for the $\ktre$ contributions

\allowdisplaybreaks{\begin{eqnarray}
\Delta \hat{r}^{(2,\ktre)}_{W} & =&\left(\frac{\alc}{4\pi s^2}\right)^2\Bigg\{\Bigg[\frac{1}{64} \mhw \left(-12 \mhw^2+49 \mhw+18\right) +\mhw \frac{4 \mhw^2-7 \mhw+6}{16 \left(\mhw-1\right)} \lnb\left(\mw^2\right) \nn\\
   &&+ \left(\frac{10-13 \mhw}{16 \left(\mhw-1\right)}\mhw^2+\frac{-2 \mhw^4+9 \mhw^3-46 \mhw+60}{32 \left(\mhw-1\right)^2} \mhw \lnb\left(\mw^2\right) \right)\lnb\left(\mh^2\right)\nn\\
   &&+\frac{2 \mhw^4-9 \mhw^3+46 \mhw-60}{64 \left(\mhw-1\right)^2} \mhw  \lnb\left(\mw^2\right)^2\nn\\
   &&+3\frac{2 \mhw^4-3 \mhw^3-4 \mhw^2+18 \mhw-20}{64 \left(\mhw-1\right)^2}\mhw \lnb\left(\mh^2\right)^2\nn\\
   &&+\Bigg(\frac{1}{8} \left(\mhw^2-3 \mhw-2\right) \mhw+\frac{1}{8}  \left(\mhw-2\right) \mhw\lnb\left(\mw^2\right)\nn\\
   &&-\frac{1}{8}  \left(\mhw-2\right) \mhw^2\lnb\left(\mh^2\right)\Bigg) B_0\left(\mw^2,\mh^2,\mw^2\right) \nn\\
   &&+\frac{1}{8} \left(\mhw-2\right) \mhw B_0\left(\mw^2,\mh^2,\mw^2\right)^2  +\frac{\mhw-2 }{8 \mw^2}\mhw S_0\left(\mw^2,\mw^2,\mh^2,\mh^2\right)\nn\\
   &&-\frac{1}{2}  \left(\mhw-1\right) \mhw^2 T_0\left(\mw^2,\mh^2,\mw^2,\mh^2\right)\nn\\
   &&+\frac{1}{4}  \left(\mhw-2\right) \mhw U_0\left(\mw^2,\mh^2,\mw^2,\mh^2,\mw^2\right)\nn\\
   &&-\frac{1}{8}
    \mhw \left(\mhw^2+\mhw-6\right) U_0\left(\mw^2,\mw^2,\mh^2,\mh^2,\mh^2\right)\nn\\
   &&+\frac{1}{16}
    \mh^2 \left(\mhw^3-12 \mhw+24\right)M_0\left(\mw^2,\mh^2 ,\mh^2 ,\mw^2,\mw^2,\mh^2 \right) \nn\\
   &&+3\frac{  -2 \mhw^3+\mhw^2+4 \mhw+24 }{64 \left(\mhw-1\right)^2}\mhw^2 \phi \left(\frac{1}{4}\right)  +3\frac{ 4 \mhw^2-41 \mhw+10 }{32 \left(\mhw-1\right)^2} \mhw  \phi \left(\frac{1}{4 \mhw}\right)\nn\\
   &&-\frac{ \mhw \left(2 \mhw^4-13 \mhw^3+18 \mhw^2+40 \mhw-128\right)}{64 \left(\mhw-1\right)^2} \phi \left(\frac{\mhw}{4}\right)\Bigg] \ktre\nn \\
   && \phantom{\frac{1}{2}}\nn\\%%%%%%%%%
   &&+   \Bigg[\frac{\left(-476 \mhw^4+2403 \mhw^3-4995 \mhw^2+1652 \mhw+120\right)}{256 \left(\mhw-4\right) \left(\mhw-1\right)} \mhw \nn\\
   &&+3 \frac{(\mhw^4-6 \mhw^3+39 \mhw^2-100 \mhw+12 }{32 \left(\mhw-4\right) \left(\mhw-1\right)^2}\mhw \lnb\left(\mw^2\right) \nn\\
   &&+9 \left(\frac{5 \mhw^4-31 \mhw^3+80 \mhw^2-84
   \mhw+48}{32 \left(\mhw-4\right) \left(\mhw-1\right)^2} \mhw^2  -\frac{27 \mhw^2}{32 \left(\mhw-1\right)^2} \lnb\left(\mw^2\right)\right)\lnb\left(\mh^2\right) \nn\\
   &&-3 \frac{7 \mhw^4-45 \mhw^3+117 \mhw^2-145 \mhw+120}{64 \left(\mhw-4\right) \left(\mhw-1\right)^2} \mhw^2 \lnb\left(\mh^2\right)^2 \nn\\
   &&+ \Bigg(\frac{1}{32} \mhw \left(\mhw^2-4 \mhw+12\right)-\frac{1}{16}  \mhw \left(\mhw^2-4 \mhw+12\right)\lnb\left(\mh^2\right)\nn\\
   &&-9\frac{\left(\mhw-2\right)^3}{32 \left(\mhw-4\right)} \mhw  B_0\left(\mh^2,\mh^2,\mh^2\right)\Bigg)B_0\left(\mw^2,\mh^2,\mw^2\right) \nn\\
   &&+ \Bigg(9 \frac{ 2 \mhw^4-13 \mhw^3+33 \mhw^2-36 \mhw+32}{64 \left(\mhw-4\right) \left(\mhw-1\right)} \mhw\nn\\
   &&+9 \frac{\mhw^4-6 \mhw^3+14 \mhw^2-8  \mhw+8 }{32 \left(\mhw-4\right) \left(\mhw-1\right)^2} \mhw \lnb\left(\mw^2\right) \nn\\
   &&-9 \frac{ \mhw^4-7 \mhw^3+19 \mhw^2-24 \mhw+20}{32 \left(\mhw-4\right) \left(\mhw-1\right)^2} \mhw^2 \lnb\left(\mh^2\right)\Bigg)B_0\left(\mh^2,\mh^2,\mh^2\right)\nn\\
   &&+9 \frac{\mhw^2-4 \mhw+8}{32 \mw^2 \left(\mhw-4\right)} \mhw S_0\left(\mw^2,\mw^2,\mh^2,\mh^2\right)\nn\\
   &&-\frac{\mhw^3-5 \mhw^2+16 \mhw-12}{4 \left(\mhw-4\right)} \mhw T_0\left(\mw^2,\mh^2,\mw^2,\mh^2\right)\nn\\
   &&+\frac{ 7 \mhw^3-38 \mhw^2+52 \mhw+24}{32 \left(\mhw-4\right)}\mhw U_0\left(\mw^2,\mw^2,\mh^2,\mh^2,\mh^2\right)\nn\\
   &&+3 \frac{7 \mhw^4-45 \mhw^3+99 \mhw^2-64 \mhw+84}{64 \left(\mhw-4\right) \left(\mhw-1\right)^2} \mhw^2 \phi \left(\frac{1}{4}\right)\nn\\
  &&+27 \frac{4 \mhw-1 }{64 \left(\mhw-1\right)^2}\mhw^2\phi \left(\frac{1}{4 \mhw}\right)\Bigg] \ktre^2\Bigg\}\,,
\label{drw}
\end{eqnarray}}

\beqn
Y^{(2,\ktre)}_{\ms} &=& \left(\frac{\alc}{4\pi s^2}\right)^2\Bigg\{\left[ f_1 \left(\frac{\mh^2}{\mw^2}\right) - \frac{1}{c^4} f_1 \left(\frac{\mh^2}{\mz^2}\right) \right] \ktre \nn \\
&&+  \left[f_2 \left(\frac{\mh^2}{\mw^2}\right) - \frac{1}{c^4} f_2 \left(\frac{\mh^2}{\mz^2}\right)\right]\ktre^2\Bigg\}\,,
\label{yms}
\eeqn
where we have defined the functions $f_1$, $f_2$ as

\begin{eqnarray}
f_1(\zeta\equiv \mh^2/m^2)& =&\frac{1}{32} \Bigg[-\left(6 \zeta^2-11 \zeta-15\right)\zeta +4  (2 \zeta-3)\zeta \lnb \left(m^2\right) +(\zeta-4) \zeta^2 \lnb \left(m^2\right)^2\nn\\
   &&-2 \bigg(10  \zeta^2+(\zeta-4) \zeta^2 \lnb \left(m^2\right) \bigg)\lnb \left(\mh^2\right) +3 \zeta^3 \lnb \left(\mh^2\right)^2\nn\\
   &&+4 \bigg(- 2 - 3 \zeta + \zeta^2+(\zeta-2) \lnb \left(m^2\right) \nn\\
   &&-(\zeta-2) \zeta \lnb \left(\mh^2\right)\bigg) \zeta  B_0\left(m^2,\mh^2,m^2\right) \nn\\
   &&+4 (\zeta-2) \zeta B_0\left(m^2,\mh^2,m^2\right)^2\nn\\
   && +4 (\zeta-2)\frac{ \zeta^2 }{\mh^2}S_0\left(m^2,m^2,\mh^2,\mh^2\right) \nn\\
   &&-16 (\zeta-1) \zeta^2 T_0\left(m^2,\mh^2,m^2,\mh^2\right)\nn\\
   && +8 (\zeta-2) \zeta U_0\left(m^2,\mh^2,m^2,\mh^2,m^2\right)\nn\\
   &&-4 \left(\zeta^2+\zeta-6\right)\zeta U_0\left(m^2,m^2,\mh^2,\mh^2,\mh^2\right)\nn\\
   && +2 \left(\zeta^3-12 \zeta+24\right) \mh^2 M_0\left(m^2,\mh^2,\mh^2,m^2,m^2,\mh^2\right)\nn\\
  &&-3 \zeta^3 \phi \left(\frac{1}{4}\right) -(\zeta-4) (\zeta-2) \zeta \phi \left(\frac{\zeta}{4}\right) \Bigg]\,,
\label{f1f}
   \end{eqnarray}
\begin{eqnarray}
f_2(\zeta\equiv \mh^2/m^2)& =&\frac{1}{128} \Bigg[-\frac{\left(238 \zeta^3-941 \zeta^2+1660 \zeta+60\right) \zeta}{\zeta-4} +\frac{12 \left(\zeta^2-4 \zeta+12\right)  \zeta}{\zeta-4}\lnb \left(m^2\right)\nn\\
   &&+36 \frac{5 \zeta^2-20 \zeta+32}{\zeta-4}\zeta^2 \lnb \left(\mh^2\right) - 6 \frac{ 7 \zeta^2-28 \zeta+36}{\zeta-4}\zeta^2 \lnb \left(\mh^2\right)^2\nn\\
   &&+36  \Bigg(\frac{ \zeta^3-5 \zeta^2+12 \zeta-16}{\zeta-4}\zeta +\frac{ \zeta^2-4 \zeta+8}{\zeta-4}\zeta\lnb \left(m^2\right)\nn\\
   &&- \frac{(\zeta-2)^2 }{\zeta-4}\zeta^2\lnb \left(\mh^2\right)-\frac{ (\zeta-2)^3 }{\zeta-4}\zeta B_0\left(m^2,\mh^2,m^2\right)\Bigg) B_0\left(\mh^2,\mh^2,\mh^2\right) \nn\\
   &&+4 \left(\zeta^2-4 \zeta+12\right) \bigg(1-2 \lnb \left(\mh^2\right)\bigg) \zeta B_0\left(m^2,\mh^2,m^2\right) \nn\\
   &&+36\frac{ \zeta^2-4 \zeta+8}{\mh^2(\zeta-4)}\zeta^2 S_0\left(m^2,m^2,\mh^2,\mh^2\right)\nn\\
   &&-32\frac{ \zeta^3-5 \zeta^2+16 \zeta-12}{\zeta-4} \zeta T_0\left(m^2,\mh^2,m^2,\mh^2\right)\nn\\
   &&+4\frac{7 \zeta^3-38 \zeta^2+52 \zeta+24 }{\zeta-4}\zeta U_0\left(m^2,m^2,\mh^2,\mh^2,\mh^2\right) \nn\\
  &&+6 \frac{ 7 \zeta^2-28 \zeta+36}{\zeta-4}\zeta^2 \phi \left(\frac{1}{4}\right) \Bigg]\,.
\label{f2f}
\end{eqnarray}
In eqs.(\ref{drw}--\ref{f2f})
\beq
\phi \left(x\right)=
      4\sqrt{\frac{x}{1-x}}\ \text{Im}(\text{Li}_2 (e^{i 2\arcsin(\sqrt{x})}))\,,
\eeq
and, following Refs.~\cite{Martin:2003qz,Martin:2005qm}, we define the
$d$-dimensional functions
\beqn
B_0(s,x,y)&=&\lim_{\epsilon\rightarrow 0}\left[B(s,x,y)-\frac{1}{\epsilon}\right]=-\int_0^1 dt \lnb[tx+(1-t)y-t(1-t)s]\,, \\
S_0(s,x,y,z)&=&\lim_{\epsilon\rightarrow 0}\left[S(s,x,y,z)+\frac{x+y+z}{2 \epsilon^2}+\frac{\frac{s}{2}-x-y-z}{2\epsilon}-\frac{A(x)+A(y)+A(z)}{\epsilon}\right]\,, \nonumber \\
\\
T_0(s,x,y,z)&=&-\frac{\partial}{\partial x}S_0(s,x,y,z)\,,\\
U_0(s,x,y,z,u)&=&\lim_{\epsilon\rightarrow 0}\left[U(s,x,y,z,u)+\frac{1}{2 \epsilon^2}-\frac{1}{2\epsilon}-\frac{B(s,x,y)}{\epsilon}\right]\,,\\
M_0(s,x,y,z,u,v)&=&\lim_{\epsilon\rightarrow 0}\left[M(s,x,y,z,u,v)\right],
\eeqn
with $d=4-2\epsilon$ and 
\beqn
A(x)&=&-i\frac{(2 \pi \mu)^{2\epsilon}}{\pi^2}\int\frac{d^d k_1 }{\Big(k_1^2-x\Big)}\,, \\
B(s,x,y)&=&-i\frac{(2 \pi \mu)^{2\epsilon}}{\pi^2}\int\frac{d^d k_1 }{\Big(k_1^2-x\Big)\Big(k_3^2-y\Big)}\,, \\
S(s,x,y,z)& =&-\left(\frac{(2 \pi \mu)^{2\epsilon}}{\pi^2}\right)^2\int \int\frac{d^d k_1 d^d k_2}{\Big(k_1^2-x\Big)\Big(k_5^2-y\Big)\Big(k_4^2-z\Big)}\,, \\
U(s,x,y,z,u)& =&- \left(\frac{(2 \pi \mu)^{2\epsilon}}{\pi^2}\right)^2 \int \int\frac{d^d k_1 d^d k_2}{\Big(k_5^2-u\Big)\Big(k_2^2-x\Big)\Big(k_3^2-z\Big)\Big(k_4^2-y\Big)}\,, \\
M(s,x,y,z,u,v) & =&- \left(\frac{(2 \pi \mu)^{2\epsilon}}{\pi^2}\right)^2 \int \int\frac{d^d k_1 d^d k_2}{\Big(k_1^2-x\Big)\Big(k_2^2-y\Big)\Big(k_3^2-z\Big)\Big(k_4^2-u\Big)\Big(k_5^2-v\Big)}\,,\nonumber \\
\eeqn
where we introduced the notation
\beq
k_3 = k_1-p\,,  \quad k_4 = k_2-p\,, \quad k_5 = k_1-k_2\,,
\eeq
with $p^2=s$. 
\end{appendletterA}

\bibliographystyle{JHEP}
\bibliography{DFG}

\end{document}